\pdfoutput=1

\documentclass{spie}  %>>> use this instead for A4 paper

\usepackage{amsmath,amsfonts,amssymb}
\usepackage{fontawesome}
\usepackage{graphicx}
\usepackage[colorlinks=true, allcolors=blue]{hyperref}
\usepackage{siunitx}
\usepackage[caption=false]{subfig}
\usepackage[utf8x]{inputenc}
\usepackage[version=4]{mhchem}

\DeclareSIUnit \Jansky {Jy}

\title{Pre-deployment Verification and Predicted Mapping Speed of MUSCAT}

\author[a]{T.~L.~R.~Brien}
\author[a]{P.~A.~R.~Ade}
\author[b,c]{P.~S.~Barry}
\author[d]{E.~Castillo-Dom\'{i}nguez}
\author[e]{D.~Ferrusca}
\author[e]{V.~G\'{o}mez-Rivera}
\author[a]{P.~Hargrave}
\author[e]{J.~L.~Hern\'{a}ndez~Rebollar}
\author[a]{A.~Hornsby}
\author[e]{D.~H.~Hughes}
\author[e]{J.~M.~J\'{a}uregui-Garc\'{i}a}
\author[f]{P.~Mauskopf}
\author[e]{D.~Murias}
\author[a]{A.~Papageorgiou}
\author[g]{E.~Pascale}
\author[e]{A.~P\'{e}rez}
\author[a]{S.~Rowe}
\author[a]{M.~W.~L.~Smith}
\author[e]{M.~Tapia}
\author[a]{C.~Tucker}
\author[e]{M.~Vel\'{a}zquez}
\author[e]{S.~Ventura}
\author[a]{S.~Doyle}

\affil[a]{School of Physics \& Astronomy, Cardiff University, The Parade, CF24 3AA, Cardiff, United Kingdom}
\affil[b]{{Argonne National Laboratory, 9700 S. Cass Ave., Lemont IL, United States of America}}
\affil[c]{University of Chicago, 5640 S. Ellis Ave., Chicago IL, United States of America}
\affil[d]{SRON-Netherlands Institute for Space Research, Landleven 12, 9747 AD Groningen, Netherlands}
\affil[e]{Instituto Nacional de Astrof\'{i}sica, \'{O}ptica y Electr\'{o}nica, Luis Enrique Erro 1, Santa Mar\'{i}a Tonantzintla 72840 Puebla, Mexico}
\affil[f]{Arizona State University, Tempe, Arizona, United States of America}
\affil[g]{Dipartimiento di Fisica, La Sapienza Universit\'{a} di Roma, Piazzale Aldo Moro 5, 00185 Roma, Italy}

\authorinfo{Corresponding author: Thomas L. R. Brien (\faicon{envelope-o} \href{mailto:tom.brien@astro.cf.ac.uk}{tom.brien@astro.cf.ac.uk} \faicon{phone} \href{tel:+442920876269}{+44 (0) 29 20876269})}

\begin{document}
\maketitle

\begin{abstract}
The Mexico-UK Submillimetre Camera for AsTronomy (MUSCAT) is  a $\SI{1.1}{\milli\metre}$ receiver consisting of 1,500 lumped-element kinetic inductance detectors (LEKIDs) for the Large Millimeter Telescope (LMT; Volcán Sierra Negra in Puebla, México). MUSCAT utilises the maximum field of view of the LMT's upgraded 50-metre primary mirror and is the first México-UK collaboration to deploy a millimetre/sub-mm receiver on the Large Millimeter Telescope. Using a simplistic simulator, we estimate a predicted mapping speed for MUSCAT by combining the measured performance of MUSCAT with the observed sky conditions at the LMT. We compare this to a previously calculated bolometric-model mapping speed and find that our mapping speed is in good agreement when this is scaled by a previously reported empirical factor. Through this simulation we show that signal contamination due to sky fluctuations can be effectively removed through the use of principle component analysis. We also give an overview of the instrument design and explain how this design allows for MUSCAT to be upgraded and act as an on-sky demonstration testbed for novel technologies after the facility-class TolTEC receiver comes online.
\end{abstract}

\section{Introduction}
The Mexico-UK Submillimetre Camera for AsTronomy (MUSCAT) is the first of a suite of new receivers scheduled to be commissioned at the Large Millimeter Telescope (LMT) on Volcán Sierra Negra in Puebla, M\'{e}xico. The principal application of MUSCAT is to carry out high spatial resolution observations with class-leading mapping speeds. The focal plane of MUSCAT has 1,500 single-colour, $1.1\mbox{-}\si{\milli\metre}$ lumped-element kinetic inductance detectors\cite{Doyle2008} (LEKIDs) operating at the background limit determined by the atmospheric loading.
\begin{figure}[hbt]
  \centering
    \subfloat[\label{fig:LMT_map}]{\includegraphics[width=0.45\textwidth]{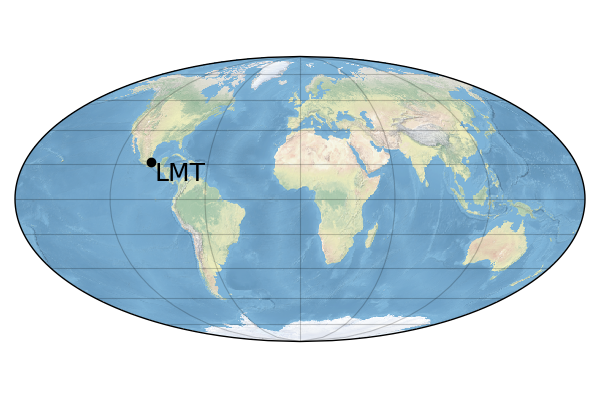}}
    \subfloat[\label{fig:h-atlas_fields}]{\includegraphics[width=0.35\textwidth]{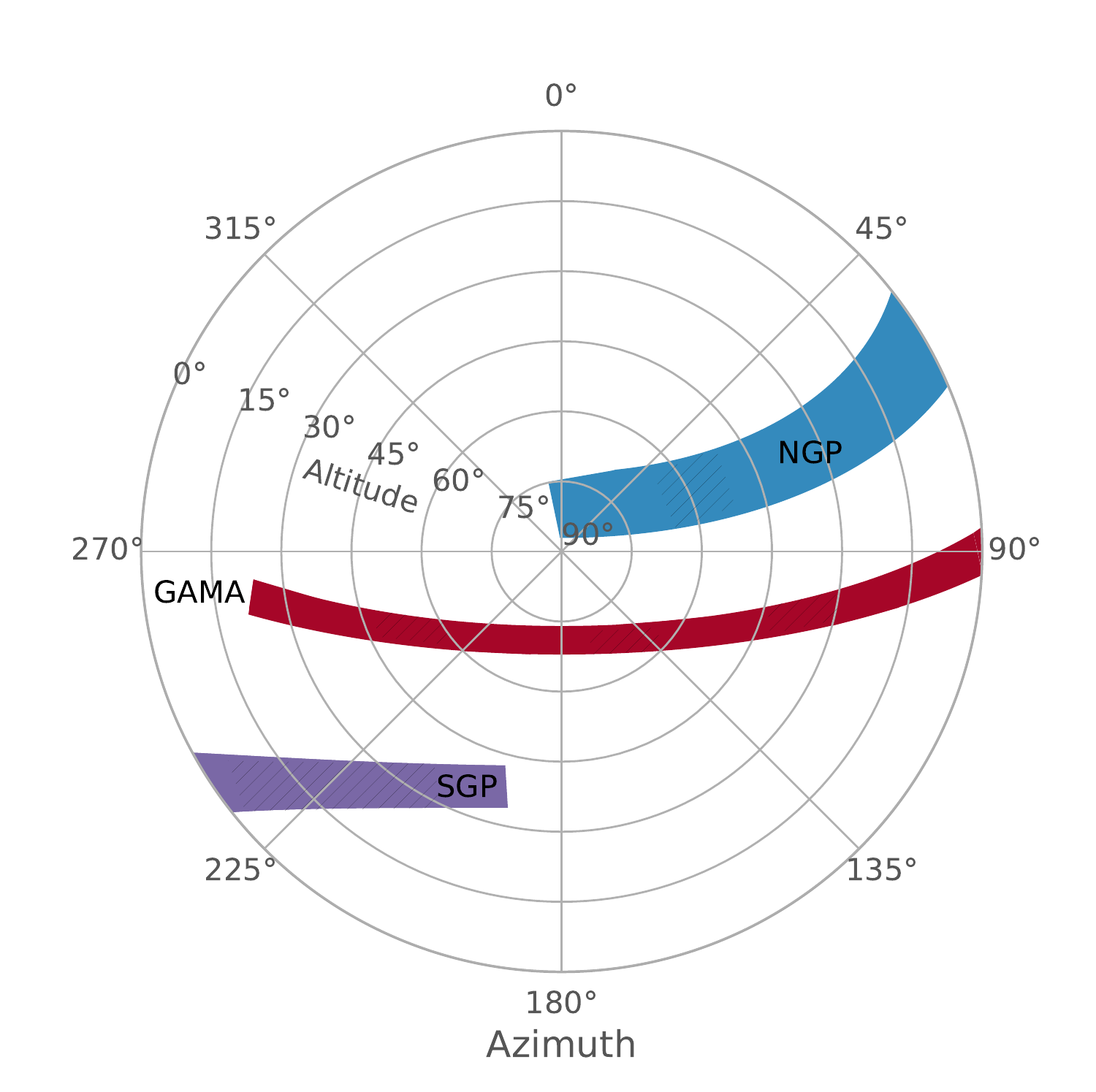}}
    \caption{\protect\subref{fig:LMT_map} Location of the Large Millimeter Telescope. Prepared with \texttt{Cartopy}\cite{Cartopy}. \protect\subref{fig:h-atlas_fields} The 5 \emph{H}-ATLAS fields as viewed from the LMT on an arbitrary night during the good weather season. The solid-shaded regions show the passage of the fields across the sky during astronomical night and the hatched areas show the instantaneous area covered by each field. In descending order of azimuth, the three GAMA fields shown are GAMA9, GAMA12 and GAMA15. Positions are calculated using the \texttt{coordinates} module of \texttt{Astropy}.\cite{astropy2.0}}
\end{figure}
\par
In its first semester on sky, MUSCAT will carry out high-resolution follow-up observations of sources identified in the \emph{Herschel}-ATLAS\cite{Valiante2016} (\emph{H}-ATLAS) survey. Of the 113,990 sources identified by \emph{H}-ATLAS in the GAMA9, GAMA12 and GAMA15 fields, only $39~\si{\percent}$ of those with a signal-to-noise ratio (SNR) of greater than four have an optical counterpart allocated.\cite{Valiante2016, Bourne2016} In the north galactic plane (NGP), only $36.4~\si{\percent}$ of the 112,155 sources meeting the same SNR threshold have optical counterparts assigned.\cite{Furlanetto2018,Valiante2016} The LMT's site on Volcán Sierra Negra ($\ang{18;59;09}~\mathrm{N}$ $\ang{97;18;53}~\mathrm{W}$) offers good coverage of both of these fields as seen in Figure~\ref{fig:h-atlas_fields} which also shows the south galactic plane (SGP). The three GAMA fields are each $12\times3~\si{\deg}$, the NGP is $15\times10~\si{\deg}$.

\par

A major limiting factor in the assignment of counterparts to faint sources detected by \emph{H}-ATLAS is the confusion noise\cite{Valiante2016, Bourne2016}. This limit is due to spatially over-lapping emission of extra-galactic sources due to the limited resolution as the wavelength increases into the far-infrared/sub-mm regime. This problem can however be greatly reduced with improved angular resolution, with the confusion limit scaling as $\left(\lambda/D\right)^{2.5}$, where $\lambda$ is the observation wavelength and $D$ is the diameter of the telescope's primary aperture.\cite{Helou1990} For the case of \emph{Herschel}, the angular resolution is defined by its $3.5~\si{\metre}$ primary mirror,\cite{Pilbratt2010} giving an angular resolution of $\ang{;;18.2}$ at a wavelength of $250~\si{\micro\metre}$ and a confusion limit of $\sim  3~\si{\milli\Jansky\per beam}$.\cite{Smith2017} The LMT and its recently upgraded $50~\si{\metre}$ primary mirror, offers an angular resolution of $\ang{;;5.5}$ allowing for substantial lowering of the confusion limit. The optical design of MUSCAT maximises the $\ang{;4;}$ diameter field of view of LMT, making MUSCAT particularly powerful for high-resolution mapping and follow-up surveys.

\section{Instrument Design}\label{sec:design}
\begin{figure}[hbt]
  \centering
    \subfloat[\label{fig:MUSCAT_xSection}]{\includegraphics[width=0.4\textwidth]{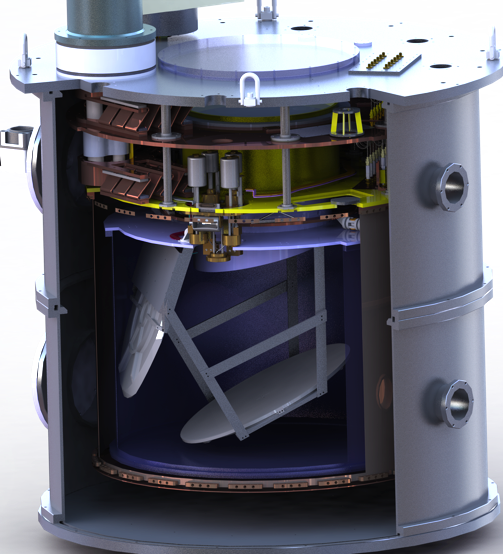}}
    \hspace{0.05\textwidth}
    \subfloat[\label{fig:MUSCAT_photo}]{\includegraphics[width=0.4\textwidth]{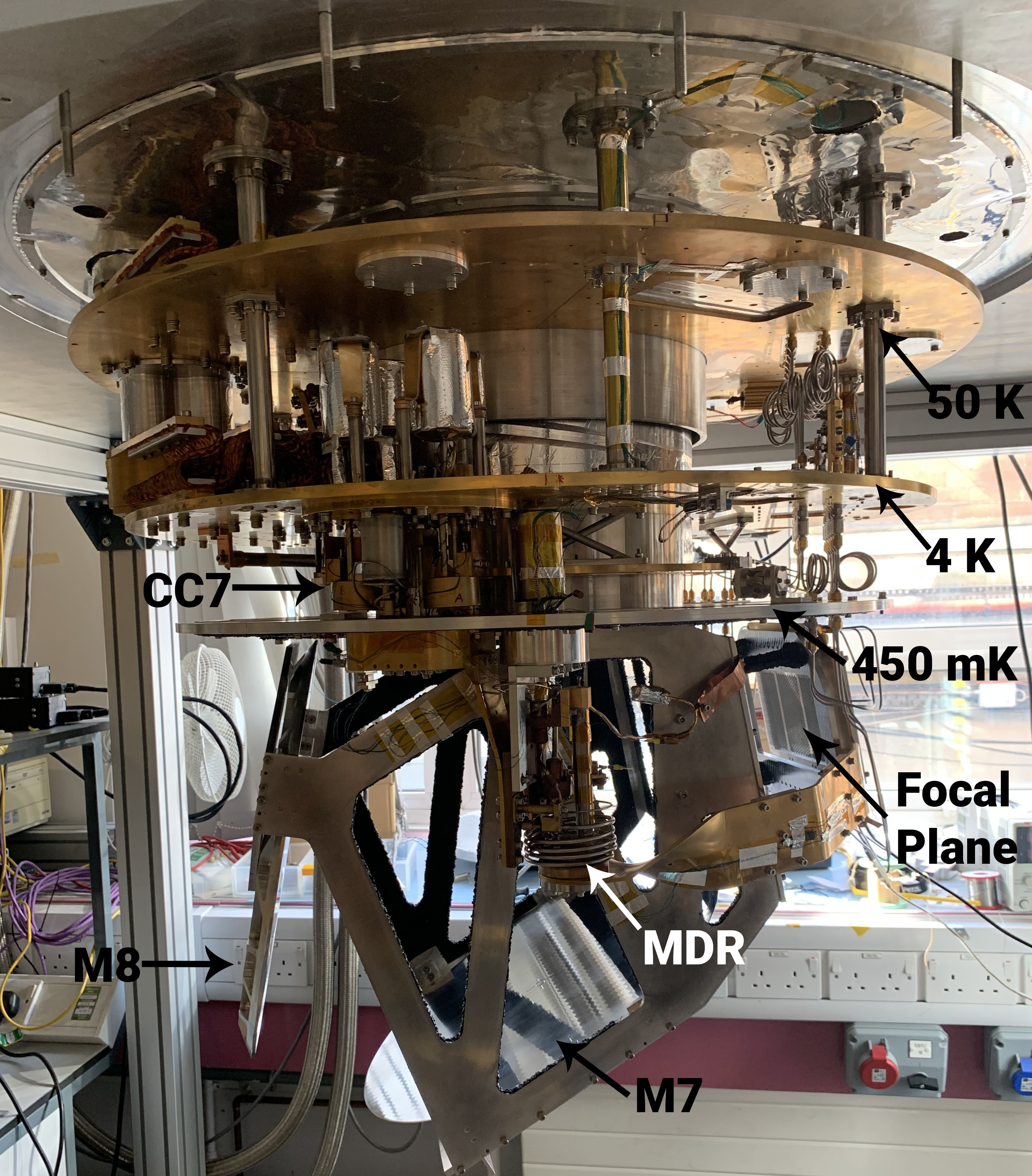}}
    \caption{\protect\subref{fig:MUSCAT_xSection} Cross section of the MUSCAT CAD model showing the nested radiation shields and the cold mirrors mounted to the $\SI{450}{\milli\kelvin}$ stage. \protect\subref{fig:MUSCAT_photo} The built system in the lab with the vacuum can and radiation shields removed, the focal plane can be seen to the right of the photo about two-thirds down.}\label{fig:cryostat}
\end{figure}

The MUSCAT receiver is based on a classic Matryoshka doll (Russian doll) cryostat, consisting of concentric radiation shields at $\SI{450}{\milli\kelvin}$, $\SI{4}{\kelvin}$, and $\SI{50}{\kelvin}$, all nested inside an outer vacuum can (OVC). Figure~\ref{fig:MUSCAT_xSection} shows a cross-sectioned CAD model of the MUSCAT cryostat and Figure~\ref{fig:MUSCAT_photo} shows the completed MUSCAT instrument during lab verification with the OVC and radiation shields removed.
\par
MUSCAT uses a fully reflective optics train which provides high-quality, diffraction-limited beams in MUSCAT's $\SI{1.1}{\milli\metre}$ band and also at $\SI{850}{\micro\metre}$ as described in Brien et al.\cite{Brien2018} In Figure~\ref{fig:cryostat} the final two mirrors (M7 and M8) can be seen mounted to the $\SI{450}{\milli\kelvin}$ plate.
\par
The design of MUSCAT's cooling system has been previously described\cite{Brien2018cryo} and utilises four different cooling systems to produce a continuous temperature, under the final loading, of $\SI{133}{\milli\kelvin}$ at the focal plane. Initial cooling of the 50- and 4-kelvin stages is performed using a Cryomech PT-420-RM which provides $\SI{2.0}{\watt}$ of cooling at $\SI{4.2}{\kelvin}$. A 1-kelvin stage is cooled by a Chase Research Cryogenics \ce{^{4}He} continuous sorption cooler (CC4); this stage does not include a radiation shield and serves to buffer colder stages from thermal loading via the mechanical supports to the $\SI{4}{\kelvin}$ stage. A further Chase Research Cryogenics continuous sorption cooler using \ce{^{3}He} evaporators in addition to \ce{^{4}He} pre-cooling stages (CC7), is used to cool the $\SI{450}{\milli\kelvin}$ stage. Finally, a miniature dilution refrigerator (MDR) is used to cool the focal plane to $\sim\SI{130}{\milli\kelvin}$. It should be noted that while the CC7 stage runs continuously at a much higher temperature than the performance reported by Klemencic et al.,\cite{Klemencic2016} this is due to operating under a substantial thermal load, approximately $\SI{650}{\micro\watt}$, which is due to thermal loading from the $\SI{4}{\kelvin}$ and $\SI{1}{\kelvin}$ stages and also from heating the \ce{^{3}He} evaporator of the miniature dilution refrigerator.
\par
The first-generation MUSCAT focal plane design contains 1,500 LEKIDs which are distributed across three $\SI{100}{\milli\metre}$ silicon wafers with two sub-arrays (250 pixels) on each wafer. Each sub-array is read out on a single RF channel (MUX ratio 250:1). The detectors have been shown\cite{Gomez2020} to operate at the background limit for source temperatures above $\SI{10}{\kelvin}$. The focal plane has been shown to have a science-grade yield of in excess of $\SI{75}{\percent}$.\cite{Tapia2020}

\subsection{Focal Plane Vibration Mitigation}
\begin{figure}[ht]
  \centering
    \subfloat[\label{fig:detecotrModule}]{\includegraphics[height=0.2\textheight]{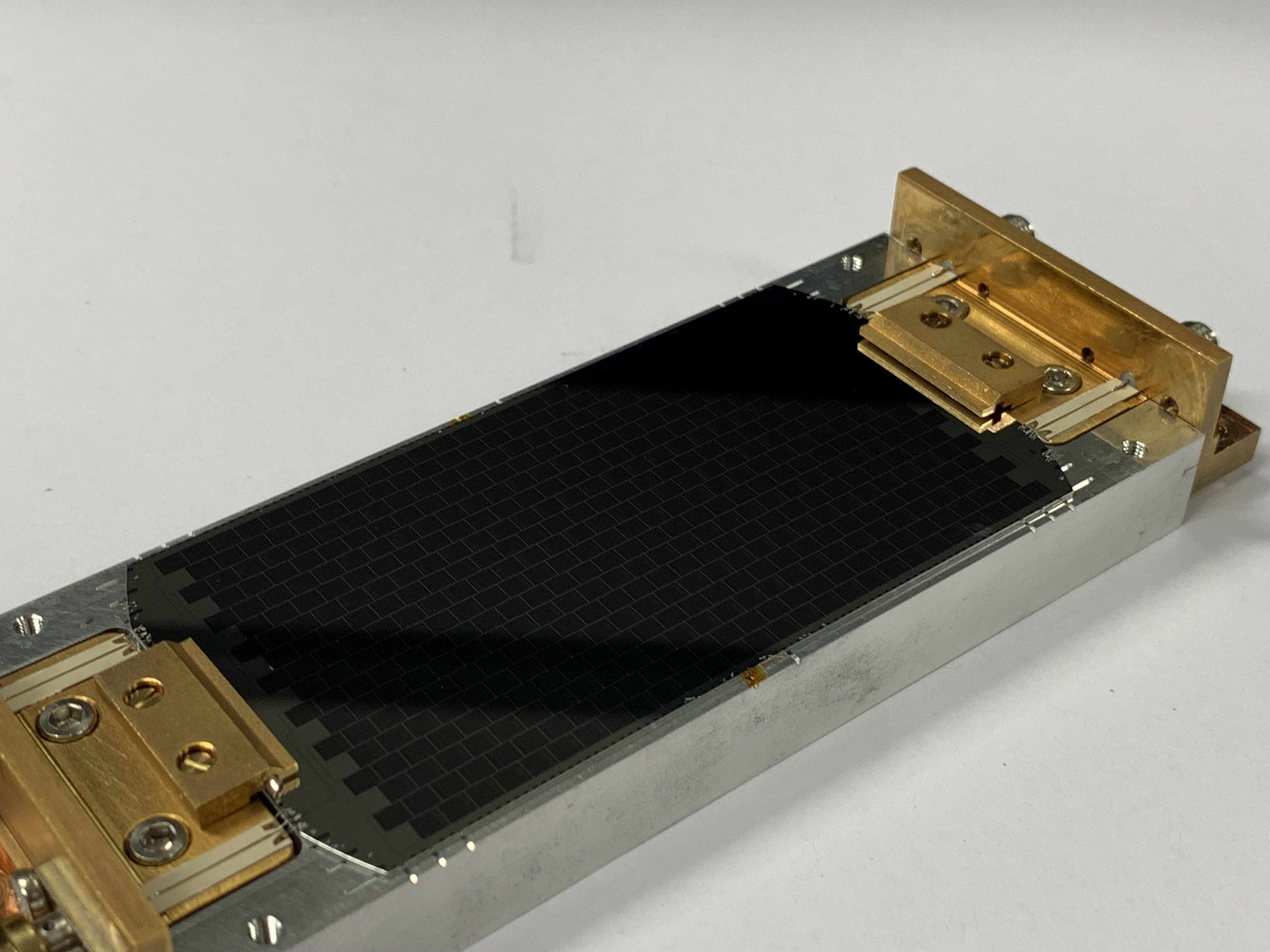}}
    \hspace{0.02\textwidth}
    \subfloat[\label{fig:pogoPin}]{\includegraphics[height=0.2\textheight]{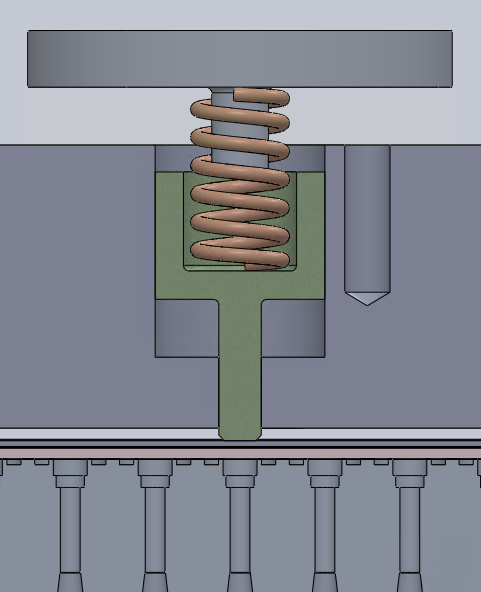}}
    \hspace{0.02\textwidth}
    \subfloat[\label{fig:vibrationPSDs}]{\includegraphics[height=0.2\textheight]{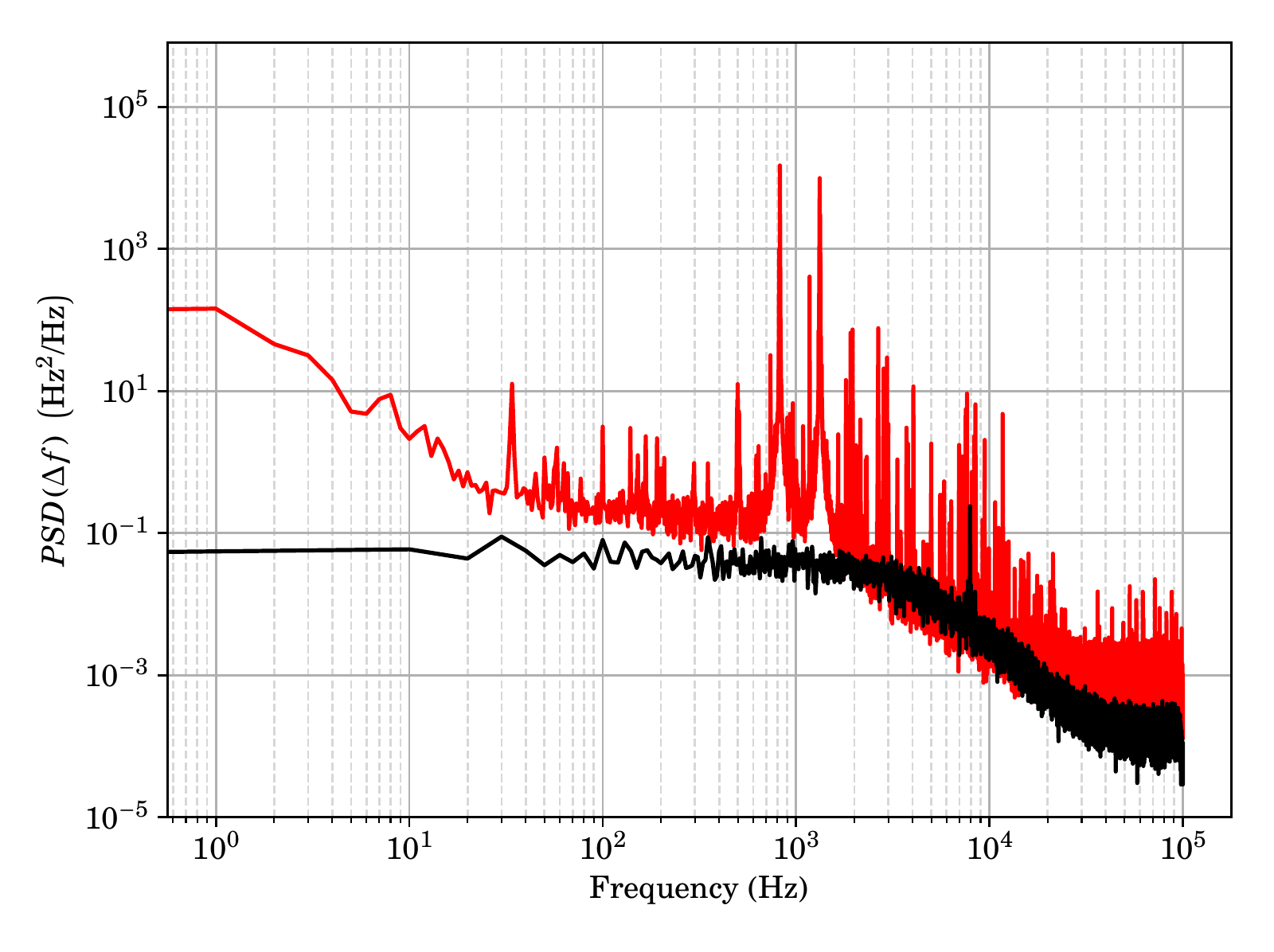}}
    \caption{\protect\subref{fig:detecotrModule} View of one of the three focal plane modules (with the back short removed). \protect\subref{fig:pogoPin} Cross sectioned CAD model of a vibration suppressing pin, shown in uncompressed state. When compressed, the top cap is secured to the back short by three equally-spaced bolts, the pin contacts directly to the detector wafer. \protect\subref{fig:vibrationPSDs} Power spectral densities without (red, background) and with (black, foreground) the vibration suppressing pins. For a complete discussion of the noise spectra, we direct readers to Tapia et al.\cite{Tapia2020}}
\end{figure}

As described in Section~\ref{sec:design}, the MUSCAT focal plane is distributed across three $\SI{100}{\milli\metre}$ silicon wafers. In order to minimise the gap between wafers, the diced wafers are only clamped along their short edges as seen in Figure~\ref{fig:detecotrModule}. Early testing of MUSCAT showed a substantial level of vibrational pickup in the noise power spectrum which is illustrated by the red trace in Figure~\ref{fig:vibrationPSDs} where a cluster of lines are seen from approximately $\SI{500}{\hertz}$ and the spectrum is dominated by $1/f$ noise below $\SI{30}{\hertz}$. To rectify this, a set of four spring-loaded pins were added to each detector module. A cross section of a single pin is shown in Figure~\ref{fig:pogoPin}. The springs were custom made from beryllium copper and designed to apply a force of $\SI{1}{\newton}$ each when compressed by $\SI{3}{\milli\metre}$. The pin itself was made from G10 glass epoxy selected to minimise the risk of damaging the detector wafer. The pin is compressed by an aluminium cap which is bolted to the detector block (also aluminium) and incorporates a short post to ensure the spring is only free to move in the compression direction.
\par
The black trace in Figure~\ref{fig:vibrationPSDs} shows the noise spectrum of the same detector measured after the spring-loaded pins were added. All of the high-frequency noise due to vibration has been removed except a single line feature at $\SI{7.5}{\kilo\hertz}$ which is well outside of the MUSCAT readout band ($\SI{500}{\hertz}$), the $1/f$ noise has also been removed with the spectrum now flat to $\sim\SI{1}{Hz}$.

\section{Instrument Platform Verification \& Performance}
\begin{figure}[b]
  \centering
    \includegraphics[width=0.95\textwidth]{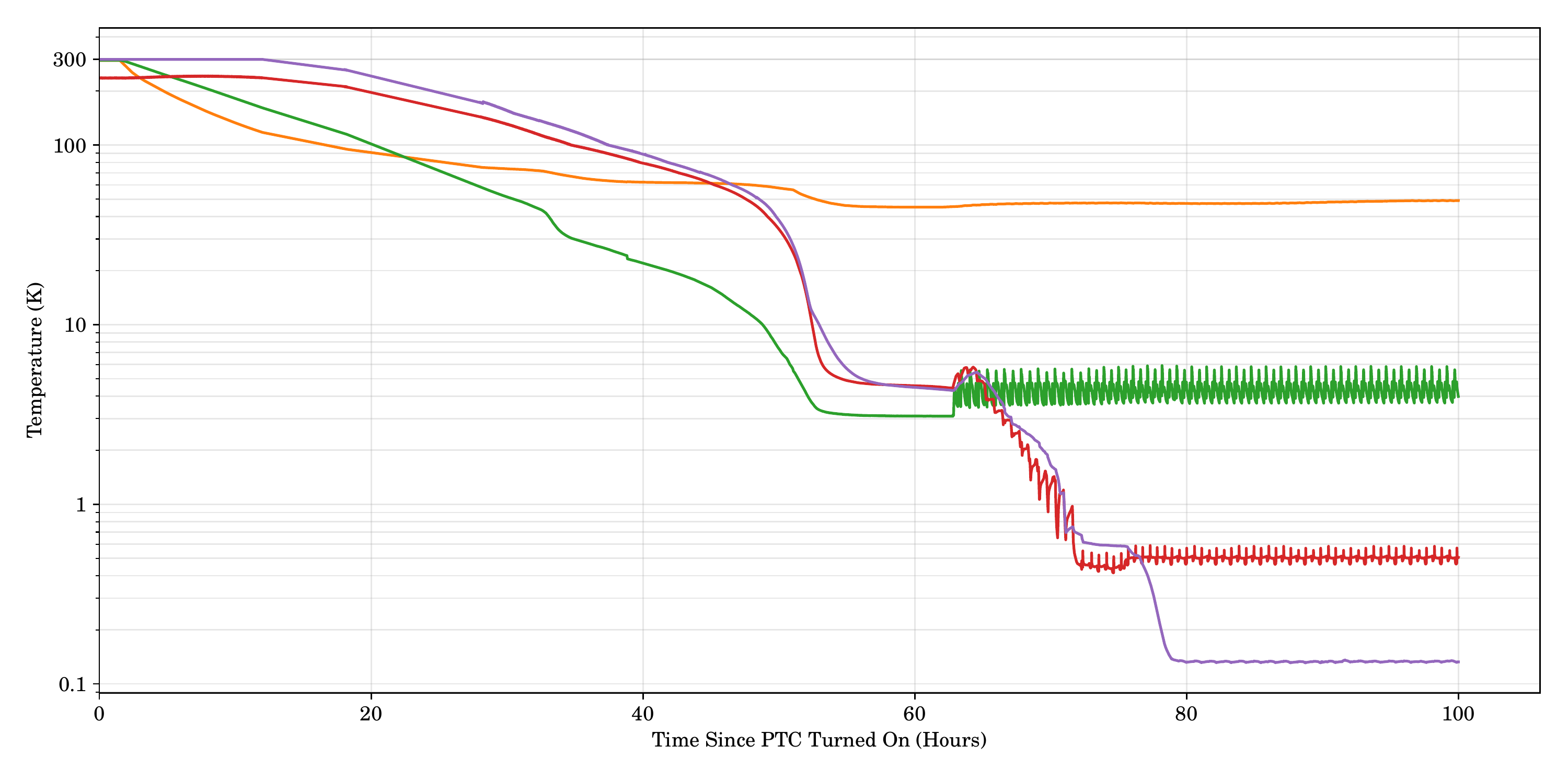}
    \caption{Cooldown of the MUSCAT instrument platform from room temperature to continuous operation temperature (\SI{130}{\milli\kelvin}). The operating temperature is achieved after approximately 80 hours. The oscillations in temperature seen in some stages after $t = \SI{60}{\hour}$ are due to the continuous recycling of the \SI{1}{\kelvin} and \SI{450}{\milli\kelvin} stage coolers. The four stages shown (in descending order of temperature at $t = \SI{100}{\hour}$) are the \SI{50}{\kelvin}, \SI{4}{\kelvin} and \SI{450}{\milli\kelvin} stages and the final (coldest) trace is the detector temperature.}\label{fig:cooldown}
\end{figure}

The cryogenic platform, as described in Section~\ref{sec:design}, has shown excellent performance in terms of stability and minimum temperature as well as usability. Figure~\ref{fig:cooldown} shows the cool down of the system from room temperature to a minimum detector temperature of $\sim\SI{130}{\milli\kelvin}$ after 80 hours. The oscillations in the $\SI{4}{\kelvin}$ and $\SI{450}{\milli\kelvin}$ stages after $t=\SI{60}{\hour}$ is due to the operation of the CC4 and CC7 systems. Circulation of \ce{^{3}He} in the MDR commences at approximately $t=\SI{75}{\hour}.$
During prolonged lab-based verification work, MUSCAT maintained a focal-plane temperature of $\left(132.6\pm0.7\right)~\si{\milli\kelvin}$ over a period of in excess of seven weeks. This shows the cooling platform offers stable temperatures on timescales in excess of those required for observatory operations. The oscillations seen in the \SI{4}{\kelvin} stage (green trace) after $t = 60~\si{\hour}$ in Figure~\ref{fig:cooldown} are a result of the heating and rapid cooling via gas-gap heat switches of the sorption pumps in the continuous coolers. The similar oscillations seen in the $\SI{450}{\milli\kelvin}$ are more extreme examples of the fluctuations already reported for this type of continuous sorption cooler when operating without a PID controlled stabilisation heater.\cite{Klemencic2016} The fluctuations seen in MUSCAT have a higher amplitude than those reported previously for this type of fridge, this is a result of the high thermal load the $\SI{450}{\milli\kelvin}$ stage is subjected to. This also causes the elevated temperature at which this stage operates compared to the ultimate performance of the continuous sorption coolers. These oscillations are not present in the focal plane (purple trace in Figure~\ref{fig:cooldown}) as the circulation rate of \ce{^{3}He} in the miniature dilution refrigerator is much slower than the oscillation period of the the fluctuations and effectively acts to dampen the fluctuation at the focal plane.

\section{Preliminary Estimation of Mapping Speed}\label{sec:mappingSpeed}

To support the development of software for science operations with MUSCAT and to help us optimise scan strategies, we have developed a simplistic mapping simulator. This simulator combines the measured detector and focal plane performance, obtained through a thorough programme of lab-based verification as reported by G\'{o}mez-Rivera et al.\cite{Gomez2020} and Tapia et al.\cite{Tapia2020} as well as the known sky conditions at the LMT site.\cite{Ferrusca2014,Zeballos2016}

\begin{figure}[htb]
  \centering
    \subfloat[\label{fig:beammap}]{\includegraphics[height=0.28\textheight]{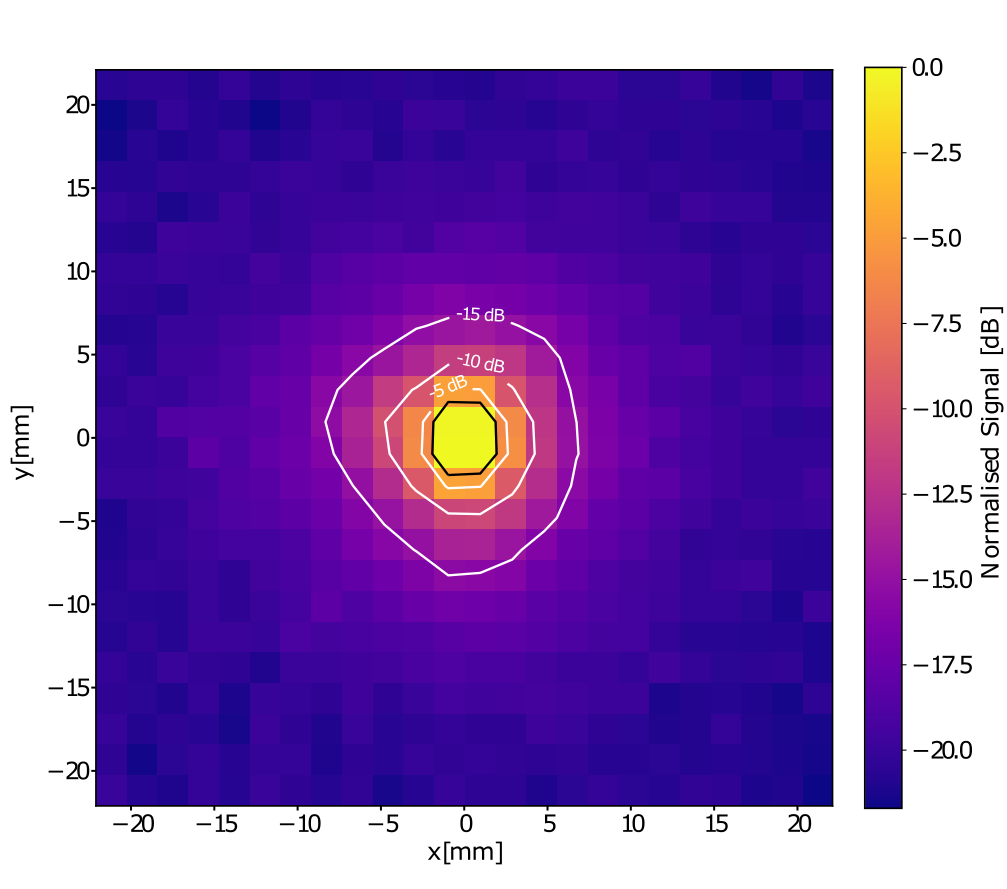}}
    \hspace{0.05\textwidth}
    \subfloat[\label{fig:NEP_PSD}]{\includegraphics[height=0.28\textheight]{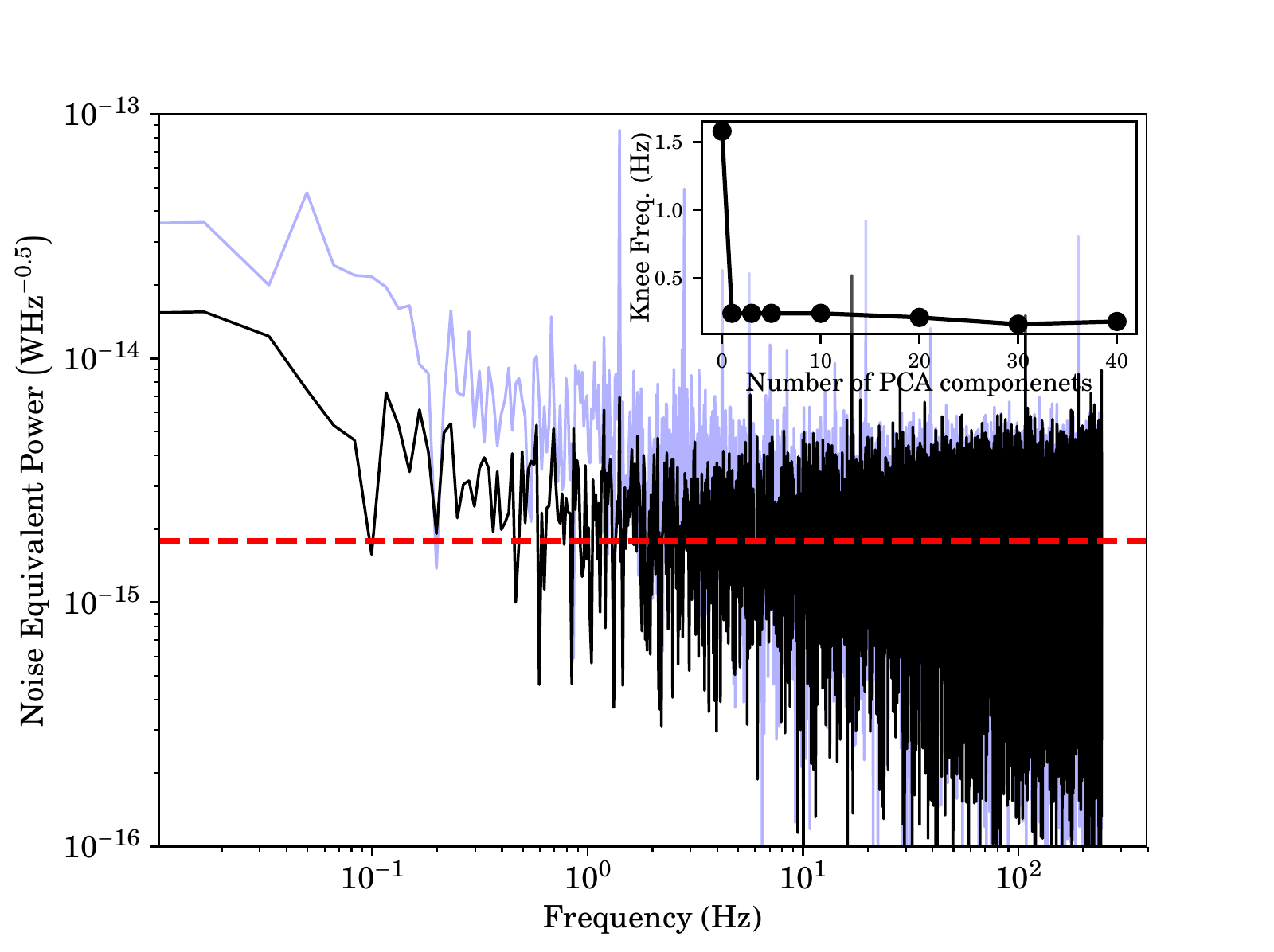}}
    \caption{Results from Tapia et al.\cite{Tapia2020} \protect\subref{fig:beammap} Stacked and centred beam maps of two sub arrays. \protect\subref{fig:NEP_PSD} Noise equivalent power (NEP) of a typical pixel in the MUSCAT focal plane. The faint blue trace in the background shows the raw NEP spectrum prior to cleaning with principal component analysis (PCA), the black trace shows the NEP after PCA cleaning. With PCA the $1/f$-noise knee frequency is typically reduced from $\SI{1.6}{\hertz}$ to $\SI{0.2}{\hertz}$. The measured NEP spectrum is in good agreement with the expected photon noise level (dashed line) of the room-temperature black body observed during these measurements. \emph{Inset:} Reduction in $1/f$-noise knee frequency with number of PCA components. (Colour online)}\label{fig:instrumentResults}
\end{figure}

Figure~\ref{fig:instrumentResults} shows some of the key results from lab-verification used to build a mapping simulator. Figure~\ref{fig:beammap} shows the combined beam maps for two sub-arrays ($\sim 500$ pixels) and Figure~\ref{fig:NEP_PSD} shows a typical noise spectrum, with and without cleaning via principal component analysis (PCA). For the purpose of estimating the mapping speed we note that the measured beams can be approximately circular down to a level of $-15~\si{\dB}$ and that the noise spectra have a $1/f$ knee frequency of $0.2~\si{\hertz}$ when cleaned using PCA. Furthermore, while Tapia et al.\cite{Tapia2020} describe the performance of MUSCAT observing a $>\SI{300}{\kelvin}$ black body, we note that the single-pixel results reported by G\'{o}mez-Rivera et al.\cite{Gomez2020} show that the $\SI{1.6}{\hertz}$ $1/f$ knee (prior to PCA cleaning) and photon-noise limited are also observed for black body temperatures in the range $10\mbox{--}70~\si{\kelvin}$ which is representative of the expected effective sky temperature at the the LMT's site.

\subsection{Description of Observation Simulation}\label{sec:SimDescription}

\begin{figure}[ht]
  \centering
    \subfloat[\label{fig:scene}]{\includegraphics[width=0.5\textwidth]{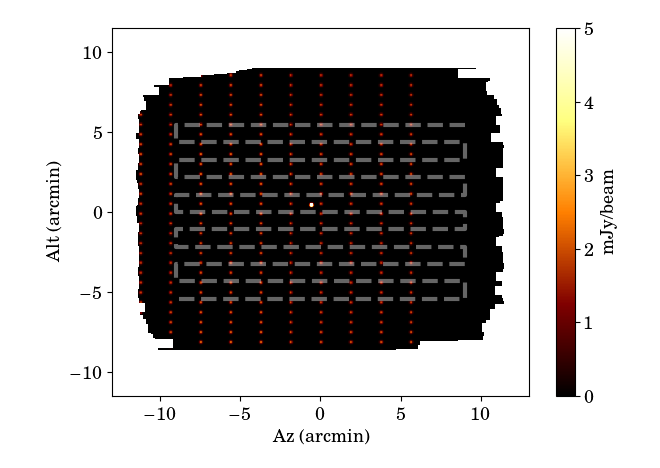}}
    \subfloat[\label{fig:sim_int}]{\includegraphics[width=0.5\textwidth]{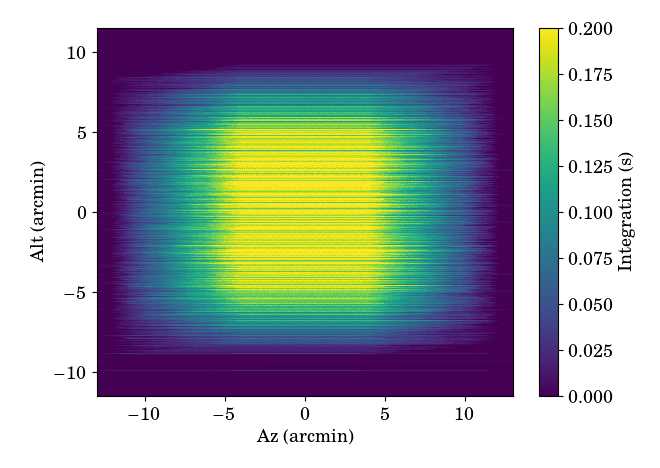}}
    \caption{\protect\subref{fig:scene} Example astronomical scene used in the simulation consisting of regularly spaced $\SI{5}{\milli\Jansky}$ sources which are produced by convolving delta functions with the beam profile shown in Figure~\ref{fig:beammap}. This figure is the recovered map from a simulation without the addition of instrument or sky noise. The transparent grey line shows the boresight path of the raster scan. \protect\subref{fig:sim_int} Integration time, per pixel, across the mapped region.}
\end{figure}

The simulated astronomical scene is a collection of $\SI{5}{\milli\Jansky}$ sources arranged in a uniform two-dimensional grid pattern against a noiseless background. These dummy point sources have been modelled by convolving a delta function with the measured beam profile (see Figure~\ref{fig:beammap}) and sampled on a grid spacing of $0.05\times\mathit{FWHM}_{\mathrm{beam}}$. Figure~\ref{fig:scene} shows this initial scene. A single brighter source is located close to the centre of the map for use as a datum.
\par
The scene is mapped using a raster-scan pattern at a rate achievable by the LMT ($\ang{;;400}\si{\per\second}$) as shown in Figure~\ref{fig:scene}. Beam offsets are taken from the lab-measured data and only the pixels found by Tapia et al. are used. This scan strategy produces a mean integration time of $\SI{0.18}{\second\per pix}$ (see Figure~\ref{fig:sim_int}) in a $\ang{;10;}\times\ang{;10;}$ square in the centre of the map and takes $\SI{29.7}{seconds}$.
\par
The response to our simulated sources is calculated using the measured responsivity curves (see G\'{o}mez-Rivera et al.\cite{Gomez2020}) and produces the clean timestream shown by the black trace in Figure~\ref{fig:timestream}. This timestream is then contaminated by a typical sky-noise spectrum (D. S\'{a}nchez-Arguelles 2019, LMT internal communication) for $\tau=0.06$ and the measured detector and system noise of MUSCAT for each individual pixel\cite{Tapia2020}, this produces the red trace in Figure~\ref{fig:timestream}. Finally, the contaminated timestream is cleaned at the sub-array level using principle component analysis (green trace in Figure~\ref{fig:timestream}).

\subsection{Simulation Assumptions}

In order to covert the noise power spectra and responsivity reported by Tapia et al.\cite{Tapia2020} and G\'{o}mez-Rivera et al.\cite{Gomez2020} into flux density units, we use the following:

\begin{align}
  F_{\si{\Jansky}} &= \frac{P\cdot10^{26}}{A_{\mathrm{tel}} \cdot \delta\nu \cdot \eta_{\mathrm{opt}}}\,, \\
  S_{\si{\hertz\per\Jansky{}}} &= S_{\si{\hertz\per\pico\watt{}}} \cdot A_{\mathrm{tel}} \cdot \delta\nu \cdot \eta_{\mathrm{opt}} \cdot 10^{-14}\,. \label{eqn:responsivity}
\end{align}

Where $F_{\si{\Jansky}}$ is the flux density in Janskys, $P$ is the optical power, $A_{\mathrm{tel}}$ is the collecting area of the telescope, $\delta\nu$ is the receiver's optical bandwidth, $\eta_{\mathrm{opt}}$ is the observation optical efficiency, and $S$ is the responsivity in the units indicated. The factor of $10^{-14}$ in Equation~\ref{eqn:responsivity} combines the conversion from Watts to Janskys and also from picowatts to Watts.
\par
We assume that the entire collecting area of the LMT is utilised, giving $A_{\mathrm{tel}} = 1960~\si{\metre^{2}}$; an optical bandwidth of $\delta\nu = 50~\si{\giga\hertz}$ according to Tapia et al.\cite{Tapia2020} We use a constant value of $\eta_{\mathrm{opt}} = 0.7$ which is dominated by the sky---the instrumental efficiency is accounted for in responsivity. As mentioned in Section~\ref{sec:SimDescription}, we assume a constant opacity of $\tau = 0.06$ during the observation, this lies within the reported interquartile range for the good-weather season at the LMT.\cite{Ferrusca2014, Zeballos2016, Brien2018} Finally, we assume that all detectors see the same random sky noise.

\subsection{Simulation Results}

\begin{figure}[t]
  \centering
    \includegraphics[width=0.8\textwidth]{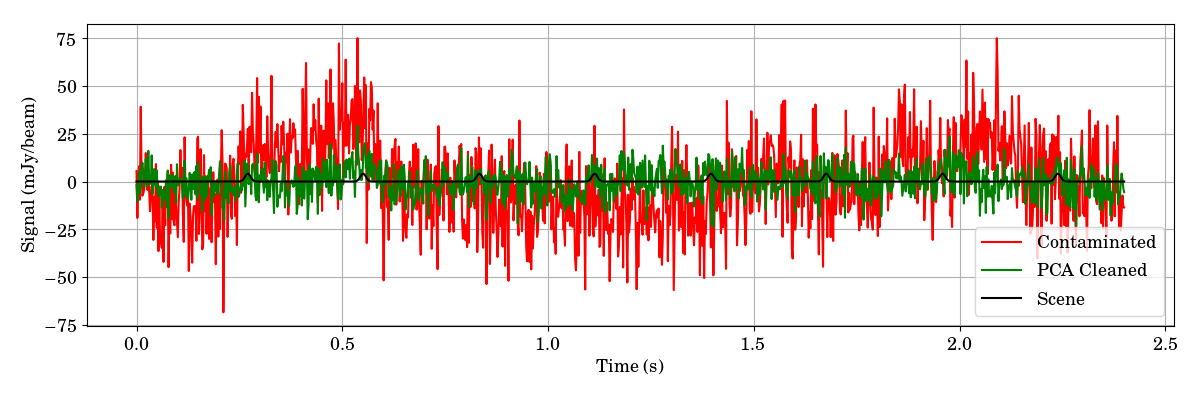}
    \caption{Example of a simulation timestream while mapping the simulated astronomical scene without noise (black trace), after contamination by sky and system noise (red trace) and after PCA cleaning (green trace).}\label{fig:timestream}
\end{figure}

\begin{figure}[ht]
  \centering
  \includegraphics[width=0.66\textwidth]{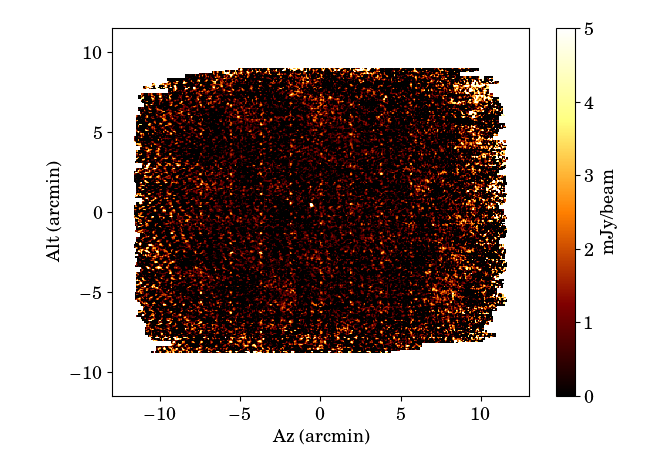}
    \caption{Reconstructed map of evenly-spaced, $\SI{5}{\milli\Jansky}$ sources after contamination by instrument and sky noise. Smoothed with a first-order Gaussian filter for clarity.}\label{fig:reconstructedImage}
\end{figure}

Figure~\ref{fig:timestream} shows the single-pixel timestream for a single azimuth scan across the scene. The flux due to the scene (black trace) is almost entirely lost in the contaminated response (red trace), however it is well recovered after PCA cleaning of the signal (green trace). Combining these timestreams for each azimuth scan with the known pointing information produces the map seen in Figure~\ref{fig:reconstructedImage} where the regular pattern of sources can be seen.

\par
To calculate the map sensitivity, we first calculate the RMS of the mapped regions known not to contain a source, correcting for the known integration time at each point (see Figure~\ref{fig:sim_int}). This produces a map sensitivity of $0.95~\si{\milli\Jansky\per\second^{0.5}}$. From this calculated mapping sensitivity, we find a simulated mapping speed of $0.63~\si{\deg^{2}\per\milli\Jansky^{2}\per\hour}$ for the scan rate of $\ang{;;400}\si{\per\second}$.

\par
This simulated mapping speed is comparable to the theoretically calculated mapping speed of $3.0~\si{\deg^{2}\per\milli\Jansky^{2}\per\hour}$ reported by Castillo-Dominguez et al.\cite{Castillo2018} using a purely bolometric model scaled by the factor $\sim 7$ found by comparing the bolometric mapping speed of the AzTEC camera to the observed performance on sky at the James Clark Maxwell (JCMT) telescope.\cite{Wilson2008} Scaling Castillo-Dominguez et al.'s mapping speed by this factor gives a mapping speed of $0.42~\si{\deg^{2}\per\milli\Jansky^{2}\per\hour}$.

\section{Status of MUSCAT}
At the time of publication, MUSCAT has completed its programme of lab-based pre-deployment verification with the results described G\'{o}mez-Rivera et al.,\cite{Gomez2020} Tapia et al.,\cite{Tapia2020} as well as those herein. Based on these results, the MUSCAT collaboration have determined that the instrument is functional and fit for purpose and MUSCAT is currently awaiting shipping to Mexico with no further lab-based studies required. Upon arrival at the LMT, MUSCAT will complete a commissioning programme to confirm that the on-sky performance is in agreement with the results of the lab-based verification. MUSCAT will then complete its 100-hour initial science run.
\begin{figure}[htb]
  \centering
  \includegraphics[width=0.75\textwidth]{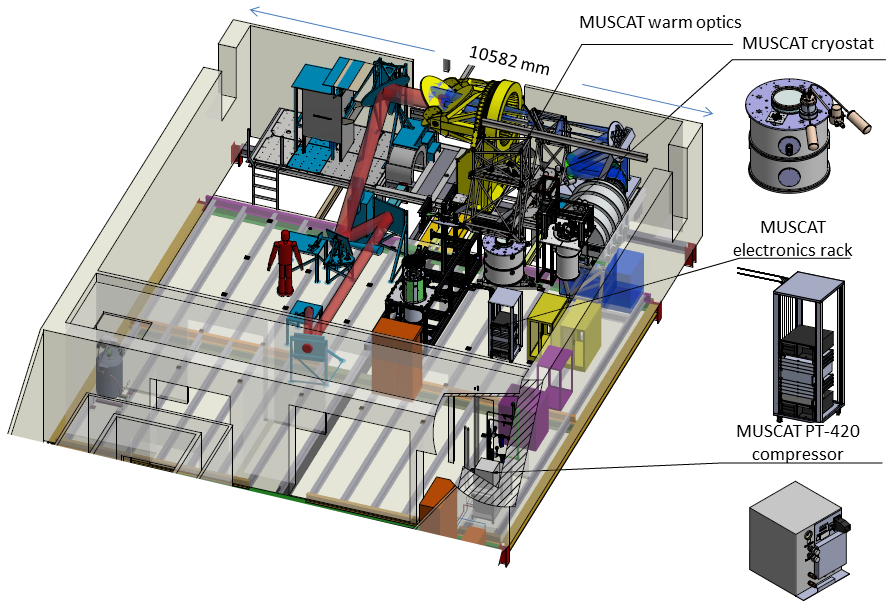}
    \caption{Model of the LMT receiver cabin showing the space that will be occupied by the MUSCAt receiver and key infrastructure. The SuperSpec and TolTEC instruments are also shown immediately to the left and right of MUSCAT respectively.}\label{fig:receiverCabin}
\end{figure}
\par
The LMT receiver cabin has been made ready for MUSCAT with the installation of warm mirror mounts and clearing of cabling and pulse tube line routes. Figure~\ref{fig:receiverCabin} shows a model of how MUSCAT will be located in the LMT receiver cabin and how MUSCAT can cohabit this space with the TolTEC\cite{toltec,Wilson2018} and SuperSpec\cite{Shirokoff2014} instruments.
\section{The Future of MUSCAT}
TolTEC is the second-generation, facility-class continuum receiver for the LMT and offers multi-colour and polarisation capabilities and will surpass the capabilities of MUSCAT in the $\SI{1.1}{\milli\metre}$ band upon its arrival. MUSCAT however is designed to provide a versatile platform for the demonstration of innovative technologies, such as a new class of mm-wave on-chip spectrometers, as well as novel KID-based readout platforms that promise significant increases in multiplexing densities. With a set of straightforward upgrades, MUSCAT is well placed to extend the capabilities offered at the LMT. As a concrete example, the design and quality of the optics make MUSCAT a platform capable of extending continuum observations at the LMT out to the $\SI{850}{\micro\metre}$ band. Through a relatively simple replacement of the focal plane and band-defining filters, MUSCAT could be upgraded to offer mapping capabilities in this band---a facility the LMT does not currently have.

\section{Conclusion}
We have shown that development of the MUSCAT instrument for the LMT has been completed, that the system provides a stable and reliable cryogenic platform and that the results of a thorough programme of lab-based verification studies show that MUSCAT is capable of carrying out high-quality astronomical observations with an anticipated mapping speed of order $\SI{0.63}{\deg^{2}\per\milli\Jansky^{2}\per\hour}$ which is in broad agreement with with a previously reported bolometric model of the the mapping speed when scaled by an accepted empirical factor. Finally, we have explained how MUSCAT can be upgraded in the future to offer further and novel astronomical capabilities at the LMT.

\section*{Acknowledgements}
We acknowledge RCUK and CONACYT through the Newton Fund (grant no. ST/P002803/1), CONACYT for supporting the fellowship for the instrument scientist (grant no. 053), Chase Research Cryogenics for the development of the sub-kelvin coolers and to XILINX Inc.\ for the donation of the FPGAs used for the ROACH2 boards.
\bibliography{bib}

\begin{thebibliography}{10}

\bibitem{Doyle2008}
S.~Doyle, P.~Mauskopf, J.~Naylon, A.~Porch, and C.~Duncombe, ``Lumped element
  kinetic inductance detectors,'' {\em Journal of Low Temperature Physics}~{\bf
  151}(1-2), pp.~530--536, 2008.

\bibitem{Cartopy}
{Met Office}, {\em Cartopy: a cartographic python library with a Matplotlib
  interface}.
\newblock Exeter, Devon, 2010 - 2015.

\bibitem{astropy2.0}
{Astropy Collaboration}, ``{The Astropy Project: Building an Open-science
  Project and Status of the v2.0 Core Package},'' {\em AJ}~{\bf 156}, p.~123,
  Sept. 2018.

\bibitem{Valiante2016}
E.~Valiante, M.~W.~L. Smith, S.~Eales, S.~J. Maddox, E.~Ibar, R.~Hopwood,
  L.~Dunne, P.~J. Cigan, S.~Dye, E.~Pascale, E.~E. Rigby, N.~Bourne,
  C.~Furlanetto, and R.~J. Ivison, ``{The Herschel--ATLAS data release 1--I.
  Maps, catalogues and number counts},'' {\em Monthly Notices of the Royal
  Astronomical Society}~{\bf 462}, pp.~3146--3179, 07 2016.

\bibitem{Bourne2016}
N.~Bourne, L.~Dunne, S.~Maddox, S.~Dye, C.~Furlanetto, C.~Hoyos, D.~Smith,
  S.~Eales, M.~W.~L. Smith, E.~Valiante, {\em et~al.}, ``The {Herschel-ATLAS}
  data release 1--ii. multi-wavelength counterparts to submillimetre sources,''
  {\em Monthly Notices of the Royal Astronomical Society}~{\bf 462}(2),
  pp.~1714--1734, 2016.

\bibitem{Furlanetto2018}
C.~Furlanetto, S.~Dye, N.~Bourne, S.~Maddox, L.~Dunne, S.~Eales, E.~Valiante,
  M.~Smith, D.~J.~B. Smith, R.~J. Ivison, {\em et~al.}, ``The second
  {Herschel-ATLAS} data release--iii. optical and near-infrared counterparts in
  the north galactic plane field,'' {\em Monthly Notices of the Royal
  Astronomical Society}~{\bf 476}(1), pp.~961--978, 2018.

\bibitem{Helou1990}
G.~{Helou} and C.~A. {Beichman}, ``{The confusion limits to the sensitivity of
  submillimeter telescopes},'' in {\em Liege International Astrophysical
  Colloquia},  B.~{Kaldeich}, ed., {\em Liege International Astrophysical
  Colloquia} {\bf 29}, pp.~117--123, Dec. 1990.

\bibitem{Pilbratt2010}
G.~Pilbratt, J.~Riedinger, T.~Passvogel, G.~Crone, D.~Doyle, U.~Gageur,
  A.~Heras, C.~Jewell, L.~Metcalfe, S.~Ott, {\em et~al.}, ``{H}erschel {S}pace
  {O}bservatory-an {ESA} facility for far-infrared and submillimetre
  astronomy,'' {\em Astronomy \& Astrophysics}~{\bf 518}, p.~L1, 2010.

\bibitem{Smith2017}
M.~W. Smith, E.~Ibar, S.~J. Maddox, E.~Valiante, L.~Dunne, S.~Eales, S.~Dye,
  C.~Furlanetto, N.~Bourne, P.~Cigan, {\em et~al.}, ``The {Herschel--ATLAS}
  data release 2, paper i. submillimeter and far-infrared images of the south
  and north galactic poles: the largest herschel survey of the extragalactic
  sky,'' {\em The Astrophysical Journal Supplement Series}~{\bf 233}(2), p.~26,
  2017.

\bibitem{Brien2018}
T.~L. Brien, P.~A. Ade, P.~S. Barry, E.~Castillo-Dom{\`\i}nguez, D.~Ferrusca,
  T.~Gascard, V.~G{\'o}mez, P.~C. Hargrave, A.~L. Hornsby, D.~Hughes, {\em
  et~al.}, ``{MUSCAT}: the mexico-uk sub-millimetre camera for astronomy,'' in
  {\em Millimeter, Submillimeter, and Far-Infrared Detectors and
  Instrumentation for Astronomy IX},   {\bf 10708}, p.~107080M, International
  Society for Optics and Photonics, 2018.

\bibitem{Brien2018cryo}
T.~L. Brien, E.~Castillo-Dominguez, S.~Chase, and S.~M. Doyle, ``A continuous
  100-{mK} helium-light cooling system for muscat on the lmt,'' {\em Journal of
  Low Temperature Physics}~{\bf 193}(5-6), pp.~805--812, 2018.

\bibitem{Klemencic2016}
G.~Klemencic, P.~Ade, S.~Chase, R.~Sudiwala, and A.~Woodcraft, ``A continuous
  dry 300 {mK} cooler for {THz} sensing applications,'' {\em Review of
  Scientific Instruments}~{\bf 87}(4), p.~045107, 2016.

\bibitem{Gomez2020}
V.~G\'{o}mez-Rivera, ``Design and characterization of the {MUSCAT} detectors,''
  {\em Millimeter, Submillimeter, and Far-Infrared Detectors and
  Instrumentation for Astronomy X} , International Society for Optics and
  Photonics, 2020.

\bibitem{Tapia2020}
M.~Tapia, ``{MUSCAT} focal plane verifcation,'' {\em Millimeter, Submillimeter,
  and Far-Infrared Detectors and Instrumentation for Astronomy X} ,
  International Society for Optics and Photonics, 2020.

\bibitem{Ferrusca2014}
D.~Ferrusca {\em et~al.}, ``Weather monitor station and 225 {GHz} radiometer
  system installed at sierra negra: the large millimeter telescope site,'' in
  {\em Ground-based and Airborne Instrumentation for Astronomy V},   {\bf
  9147}, p.~914730, International Society for Optics and Photonics, 2014.

\bibitem{Zeballos2016}
M.~Zeballos, D.~Ferrusca, D.~Hughes, {\em et~al.}, ``Reporting the first 3
  years of 225-{GHz} opacity measurements at the site of the large millimeter
  telescope alfonso serrano,'' in {\em Ground-based and Airborne Telescopes
  VI},   {\bf 9906}, p.~99064U, International Society for Optics and Photonics,
  2016.

\bibitem{Castillo2018}
E.~Castillo-Dominguez, P.~Ade, P.~Barry, T.~Brien, S.~Doyle, D.~Ferrusca,
  V.~Gomez-Rivera, P.~Hargrave, A.~Hornsby, D.~Hughes, {\em et~al.},
  ``Mexico-uk sub-millimeter camera for astronomy,'' {\em Journal of Low
  Temperature Physics}~{\bf 193}(5-6), pp.~1010--1015, 2018.

\bibitem{Wilson2008}
G.~Wilson, J.~Austermann, T.~Perera, K.~Scott, P.~A. Ade, J.~Bock, J.~Glenn,
  S.~Golwala, S.~Kim, Y.~Kang, {\em et~al.}, ``The {AzTEC} mm-wavelength
  camera,'' {\em Monthly Notices of the Royal Astronomical Society}~{\bf
  386}(2), pp.~807--818, 2008.

\bibitem{toltec}
{The TolTEC Collaboration}, ``{T}he {TolTEC} {C}amera.'' Available at
  \url{https://toltec.astro.umass.edu/} (2020/11/30).

\bibitem{Wilson2018}
G.~W. Wilson, P.~Ade, I.~Aretxaga, J.~E. Austermann, J.~Bardin, P.~Barry,
  J.~Beall, M.~Berthoud, A.~Braeley, S.~A. Bryan, {\em et~al.}, ``{The TolTEC
  project: a millimeter wavelength imaging polarimeter (Conference
  Presentation)},'' in {\em Millimeter, Submillimeter, and Far-Infrared
  Detectors and Instrumentation for Astronomy IX},   {\bf 10708}, p.~107080I,
  International Society for Optics and Photonics, 2018.

\bibitem{Shirokoff2014}
E.~Shirokoff, P.~Barry, C.~Bradford, G.~Chattopadhyay, P.~Day, S.~Doyle,
  S.~Hailey-Dunsheath, M.~Hollister, A.~Kov{\'a}cs, H.~Leduc, {\em et~al.},
  ``Design and performance of {S}uper{S}pec: an on-chip, kid-based,
  mm-wavelength spectrometer,'' {\em Journal of Low Temperature Physics}~{\bf
  176}(5-6), pp.~657--662, 2014.

\end{thebibliography}
\bibliographystyle{spiebib}
\end{document}